# Spin relaxation: From 2D to 1D

Alexander W. Holleitner

Walter Schottky Institut and Physik-Department, Technische Universität München,

Am Coulombwall 3, 85748 Garching, Germany

Abstract: In inversion asymmetric semiconductors, spin-orbit interactions give rise to very effective relaxation mechanisms of the electron spin. Recent work, based on the dimensionally constrained D'yakonov Perel' mechanism, describes increasing electron-spin relaxation times for two-dimensional conducting layers with decreasing channel width. The slow-down of the spin relaxation can be understood as a precursor of the one-dimensional limit.

Article: Semiconductor spintronics seeks to gain extra functionality compared to conventional electronics by exploiting the carrier spin degree of freedom [1,2] (Fig. 1). For a potential processing scheme which combines quantum mechanical and classical information, it is of particular interest to manipulate and to control carrier spin dynamics in non-magnetic materials by utilizing the spin-orbit interaction [3-6]. Datta and Das proposed the concept of a spin-polarized field effect transistor in narrow band gap semiconductors such as InGaAs [7]. On the one hand, the structure inversion asymmetry in these materials allows controlling the electron spin precession by an electric field [8-14]. On the other hand, the range of operation of the spin transistor is limited by the spin relaxation mechanism induced by the spin-orbit interaction [15-17]. Here, we describe the possibility of an effective suppression of the spin relaxation in the two-dimensional conducting channel of a spin transistor [18-21]. As a precursor of the one-dimensional limit, long spin relaxation times are expected when the channel width $w$ is smaller than the electronic mean free path $l_e$ ($w < l_e$). As a result, spin relaxation mechanisms, which are intrinsic to the two-dimensional electron systems (2DES), can be suppressed very efficiently [22].

In inversion asymmetric heterostructures, the spin-orbit interaction is determined by the interplay between the bulk inversion asymmetry (BIA) and the structural inversion asymmetry (SIA) and the interface asymmetry (IA) [23-26]. For zinc-blende crystals, the corresponding effective mass Hamiltionan can be written as [27]

(1) $$H = \tilde{\eta}[\sigma_x k_x (k_y^2 - k_z^2) + cycl.perm.],$$

where $\sigma_{x;y;z}$ are the Pauli matrices, and $\tilde{\eta}$ is a constant which reflects the strength of the spin-orbit interaction in the conduction band. In the case of a quasi two-dimensional quantum well, the wave vector $k_z$ along the growth direction is typically much larger than the two in-plane components $k_x$ and $k_y$. To describe the spin-orbit interactions in a semiconductor quantum well correctly, energy terms with higher orders in $k$ need to be considered [28]. However, the energy terms which are linear in $k$ are sufficient for the discussion below. Usually two Hamiltonians linear in $k$ are assumed to describe the spin-orbit interactions in a quantum well. The corresponding Hamiltonian, which refers to the BIA in a quantum well, was first described by Dresselhaus [29]

(2) $$H_{BIA} = \breve{\eta}[\sigma_x k_x + \sigma_y k_y].$$

In principle, the Dresselhaus Hamiltonian is derived from (1), by considering only terms with $k_z^2$ [19]. In asymmetric quantum wells and deformed bulk systems, the structure inversion asymmetry gives rise to the Bychkov-Rashba spin orbit coupling [30-32]

(3) $$H_{SIA} = \hat{\eta}[\sigma_x k_y - \sigma_y k_x] = \hat{\eta}\sigma[\mathbf{k} \times \mathbf{z}]$$

If an angular frequency - is defined as $\mathbf{\Omega}(\mathbf{k}) \equiv \alpha[\mathbf{v}(\mathbf{k}) \times \mathbf{z}]$ with $\mathbf{v}(\mathbf{k}) = \mathbf{\Omega}(\mathbf{k}) \equiv \hbar\mathbf{k}/m^*]$ the electron velocity, this Hamiltonian can be rewritten as

(4) $$H_{SIA} = (\hbar/2) \cdot \sigma \cdot \mathbf{\Omega}(\mathbf{k})$$

In this way, the energy splitting for electrons with opposite spin directions can be regarded as an effective magnetic field $B_{eff}(\mathbf{k})$, which depends on the wave vector of the electron. During a ballistic flight, the spin of an electron will precess with the angular frequency $\Omega(k)$, while $1/\alpha$ defines the length at which the spin of the electron rotates by an angle of $\pi$. Figure 2(b) depicts the situation in which an electric field is applied across an asymmetric quantum well. The resulting SIA of the quantum well gives rise to the effective magnetic field which is oriented in-plane. For asymmetric GaAs/Al$_x$Ga$_{1-x}$As quantum wells with an electron concentration of $10^{12}$ $cm^{-2}$, the spin splitting in the conduction band is measured to be on the order of 0.2 to 0.3 meV at the Fermi energy, corresponding to $\alpha$ = 6-9 × $10^4$ $cm^{-1}$ [19,33]. The orientation of the effective magnetic fields due to the BIA and the SIA could recently be characterized by optoelectronic measurements, in accordance with (2) and (3) [12].

Based upon the voltage tunable Rashba Hamiltonian of (3), Datta and Das proposed the concept of a spin-polarized field effect transistor in narrow band gap semiconductors such as InGaAs [7,34]. In such a spin transistor, the source and drain contacts are made out of ferromagnetic materials. As demonstrated by recent experiments [35-37], such contacts inject and collect preferentially spin polarized electrons; they act as spin polarizers and detectors. The injected spin polarized carriers precess in transit through the 2DES due to the Rashba effect. The transverse electric field can be tuned by applying a voltage to the Schottky-gate on top of the heterostructure. If the magnetization in the drain and source contacts are parallel to each other, and the carriers perform an integral number of precessions, then the device conductance is high. In the case of opposite orientation for the electron spin and the drain magnetization, the device conductance is minimized. In addition to the gate voltage, the transconductance can be also controlled via an external magnetic field. Rather than a new electronic device, the spin transistor initiated many studies on spin-polarized transport and spin-injection phenomena in semiconductor/ferromagnetic junctions [8-12, 14-17]. The spin transistor relies upon controlling the precession of the electron spin in the conducting channel due to the influence of the Rashba term. If an electron is being scattered in multiple events, the orientation of $\Omega(k)$ changes randomly. Since the elementary rotations do not commute in the plane of a quantum well (neither in the case of a three-dimensional system), the final spin orientation varies for an electron being scattered along two different trajectories from one point in space to another. Considering an ensemble of trajectories, the total spin polarization is randomized after a certain number of scattering events. The corresponding spin relaxation mechanism is named after D'yakonov and Perel' [38]. In the motional narrowing regime, i.e. the spin rotation angle is small during the ballistic flight of an electron ($\tau_P \Omega < 1$), the spin relaxation time $\tau_S$ is inversely proportional to the momentum scattering time $\tau_P$, i.e. $\tau_S^{-1} \propto \tau_P \Omega^2(k)$ [39]. In Fig. 3(a), the spin precession is sketched for an electron moving along a certain trajectory. Every time the electron is scattered, the orientation of $\Omega$ changes. Another source for spin relaxation in spin-orbit materials is given by the fact, that spin relaxation due to momentum relaxation is also possible directly through the spin-orbit coupling [40]. Generally, the Bloch states are not eigenstates of the Pauli matrixes. Thus, a lattice-induced spin-orbit interaction, e.g. influenced by phonons, can directly couple the (Pauli) spin up states to the spin down states [41]. Similarly, scattering at impurities and boundaries of the electron system can lead to instantaneous spin flip events (Fig. 3(b)) [42]. A further spin-relative relaxation mechanism was suggested by Bir, Aronov and Pikus [43]. A simultaneous spin flip of electron and hole spin, which is intermediated by the electron hole exchange interaction, can induce an effective spin relaxation of the electron spin polarization. Last but not least, the hyperfine interaction can also cause a spin flip between electrons in the conduction band and the nuclear magnetic moments of the crystal. The latter is of particular relevance for heterostructures with a nuclear magnetic moment, such as GaAs (see ref.[17] for a detailed review on spin relaxation processes in semiconductors).

In two and three dimensions the elementary rotations do not commute. As a result, scattering processes tend to randomize the spin polarization. Bournel et al. were the first to consider a 2DES with a finite width as the conducting channel of a spin transistor [18]. They studied an asymmetric quantum well made out of $In_{0,53}Ga_{0,47}As$; a system which is well known for its spin-orbit interactions due to SIA [8]. As a precursor of the one-dimensional limit, the results suggest long spin relaxation times, when the channel width $w$ is comparable to the electronic mean free path $l_e$ ($w \leq l_e$). Bournel et al. used a Monte-Carlo transport model which is based upon the collective motion of individual particles. In between instantaneous scattering events, e.g. with phonons and alloy impurities, the ballistic flight of an electron is only determined by the influence of the Rashba term. For their simulation, the authors assumed specular reflection at the boundaries of the conducting channel. At the same time, the two spin populations, injected by the source contact, are supposed to be non-intermixing. The latter condition means that other spin relaxation mechanisms than the D'yakonov Perel' mechanism, such as the one by the Elliott-Yafet, are neglected. As a result, their simulations suggest almost negligible spin relaxation in the conducting channel of an $In_{0,53}Ga_{0,47}As$ spin transistor for widths in the order of 100 *nm* and electric fields smaller than 200 *kV/cm* [18].

In a similar Monte-Carlo transport study, Kiselev and Kim identified different regimes of the spin relaxation in a 2DES with a finite width [19]. The authors assume in their work that the D'yaknov Perel' mechanism is the dominant spin relaxation mechanism in a 2DES, in which the spin-orbit interaction is dominated by the SIA term. For a very large spin splitting constant, i.e. $l_e\alpha \geq 1$, the assumptions for motional narrowing are not fulfilled anymore (see above section for the definition of $\alpha$). Here, scattering events randomize the spin information, since the precession angle of the electron spin is large during the ballistic flight of the electron ($\tau_P \Omega > 1$). In turn, spin relaxation time is proportional to the momentum scattering time $\tau_S \sim \tau_P$. In the motional narrowing regime, where $l_e\alpha < 1$, the spin relaxation time is given by $\tau_S^{2D} \sim \tau_P (l_e\alpha)^{-2}$. The authors claim that if the width *w* of the conducting channel is reduced, a transition occurs at $w\alpha \sim 1$. Kiselev and Kim find that this transition can occur at widths which are ten times larger than the mean free path. For $w\alpha < 1$, the width of the channel acts as an effective mean free path of the system. In this quasi one-dimensional regime, the spin relaxation is suppressed very efficiently as $\tau_S \sim \tau_S^{2D} (w\alpha)^{-2}$.

Another approach to the problem was given by Mal'shukov and Chao, by solving the spin diffusion equation for electrons in the conducting channel [20]. Their solutions showed that there are certain spin diffusion modes with slow spin relaxation rates, if the conducting channel of the spin transistor has got a finite width. In their model, they assume that the electron motion perpendicular to the channel can be described semi-classically. On the one hand, the width $w$ needs to be much shorter than the spin precession length $1/\alpha$, and on the other hand, it needs to be larger than the Fermi wavelength. Similar to ref. [18,19], Mal'shukov and Chao neglect the BIA, because it is assumed that in narrow gap semiconductors, the spin dynamics are dominated by the SIA. As a main result, the spin relaxation of the waveguide modes is slowed down as $\tau_S \sim (1/w\alpha)^2$, in accordance with the simulation of ref. [19]. In addition, the authors point out that the BIA term, being cubic in $k$, gives rise to an additional spin relaxation which is independent of the channel width $w$. Thus, the slow-down and finally the suppression of the spin relaxation is limited by this cubic BIA term. In a recent publication, Winkler calculated the magnitude and the orientation of $B_{eff}$ for an infinite two-dimensional InGaAs quantum well by considering higher order terms of the electron wavevector [28]. Combining both results for semi-classical wires, one expects an anisotropy of the spin relaxation times for electrons moving into different crystallographic directions.

Pareek and Bruno used a recursive Green function method, which takes into account the quantum effects at the single particle level [21]. In the case of motional narrowing, the authors again see an enhancement of the spin polarization, if the electrons are confined to a channel width of the order of the mean free path. However, they can not corroborate the result predicted by the real-space Monte Carlo simulations, that this enhancement should be detectable even for widths ten times the mean free path. In the quasi-ballistic regime, i.e. $l_e\alpha \gg 1$, Pareek and Bruno find that the spin diffusion length can be even longer than the spin precession length. At the same time, the spin relaxation rates depend on the initial orientation of the electron spin. As a result, the largest current modulation can be obtained, if the electron spin is injected perpendicular to the plane of the quantum well of the spin transistor.

Generally speaking, all theory work, which covers the effect of the transition from 2D to 1D on the spin relaxation, predicts increasing spin relaxation times as a result of a dimensionally constrained D'yakonov Perel' mechanism due to the SIA term. However, as Mal'shukov et al. pointed out, other mechanisms can limit the eventual suppression of the spin relaxation, e.g., the BIA induces a component of the D'yakonov-Perel' spin relaxation which is independent of the channel width [20]. We would like to note that in a first experimental study, an effective slowing of the Dyakonov Perel spin relaxation mechanism has been observed in conducting channels of n-InGaAs quantum wires [22, 44]. The results are consistent with a dimensionally-constrained D'yakonov-Perel' mechanism. For the narrowest wires with only a few hundreds of nanometers width, the authors find that an interplay between the spin diffusion length and the wire width determines the spin dynamics, and that channels along the crystallographic directions [100] and [010] show longer spin relaxation times than channels along [110] and [-110]. The anisotropy can be interpreted such that the cubic spin-orbit coupling terms due to BIA start to dominate the spin relaxation in the narrowest channels [24, 25, 26, 28]. We would like to further point out that a similar dimensional crossover has been observed by means of a weak antilocalization analysis of magnetotransport studies on InAs channels [45, 46, 47, 48, 49,], and by means of Shubnikov-de Haas oscillations in narrow InAs-based heterostructures [50]. Based upon these experimental results, recent theory work substiantates the interpretation that the dimensional crossover can be understood in terms of an interplay between the channel width, the spin precession length, intersubband scattering, and the effect of spin scattering at the boundaries of the channels [51, 52, 53, 54, 55, 56, 57].

To summarize, in inversion asymmetric semiconductors, the precession of the electron spin can be influenced via the spin-orbit interaction. At the same time, spin relaxation mechanisms based on the spin-orbit coupling limit the range of control on the spin dynamics. Particularly in 3D and 2D electron systems, momentum scattering gives rise to a spin relaxation mechanism which was proposed by D'yaknov and Perel'. We describe recent theoretical results, which suggest that the D'yakonov Perel' mechanism can be suppressed by decreasing the width of a two-dimensional electron system down to the electronic mean free path. In such quasi one-dimensional systems long spin relaxation times are expected.

We would like to thank D.D. Awschalom, H. Knotz, F. Meier, V. Sih, R.C. Myers, A.C. Gossard, V. Khrapey, J.P. Kotthaus for discussions and support. We gratefully acknowledge financial support from the Center for NanoScience (CeNS) in Munich, Germany, the ONR, DARPA, and the California NanoSystems Institute (CNSI) at the University of Santa Barbara, USA, the DFG Project No. HO-3324/4, and the German excellence initiative via ''Nanosystems Initiative Munich (NIM)'' and ''LMUexcellent.''

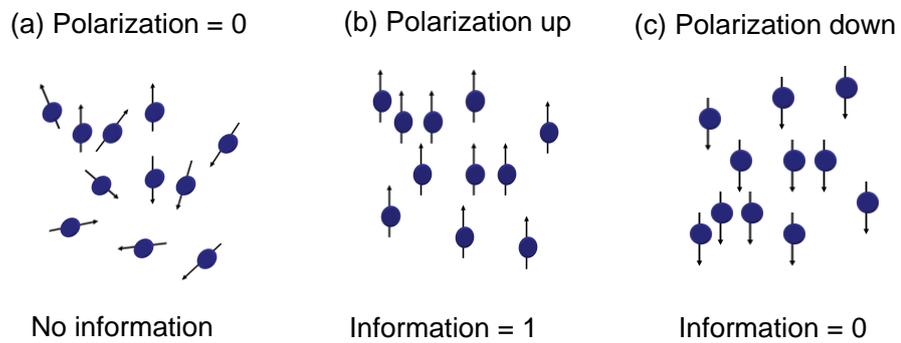

Fig. 1. In semiconductor spintronics, information is encoded via the spin polarization of the conduction electrons. (a) If the distribution of the electron spin is in equilibrium, no information is encoded. A non-equilibrium spin polarization can carry the information (b) one or (c) zero.

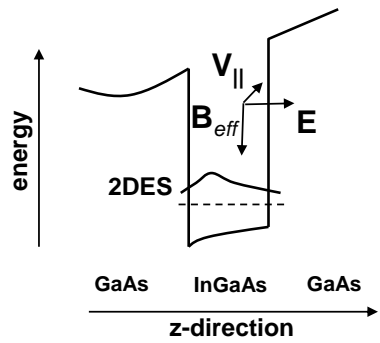

Fig. 2. Schematics of the conduction band of a heterostructure containing an asymmetric quantum well. In the example, a two-dimensional electron system (2DES) is formed in an InGaAs layer in between two GaAs layers. If an electric field E is applied along the growth direction of the heterostructure, the resulting spin-orbit interaction induces an effective magnetic field $B_{eff}(\mathbf{k})$ which is orientated in the plane of the quantum well and perpendicular to the velocity $v_{//}(\mathbf{k})$ of the electron.

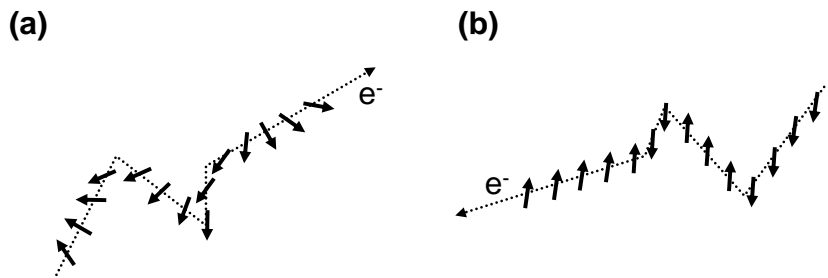

Fig. 3. Schematics of electron trajectories (dotted lines) in the case of (a) the D'yakonov Perel'mechanism and (b) the Elliot-Yafet mechanism. The orientation of the spin eigenfunction is presented with black arrows.